# Enhancement of Mid-/High-Z Impurity Transport by Continuous Li-granule Dropping in a Stellarator Plasma

Daniel Medina-Roque,[1,*] Isabel García-Cortés,[1] Naoki Tamura,[2] Kieran J. McCarthy,[1] Federico Nespoli,[3] Kenji Tanaka,[4,6] Mamoru Shoji,[4,5] Suguru Masuzaki,[4,5] Hisamichi Funaba,[4] Chihiro Suzuki,[4,5] Albert Mollén,[7] Robert Lunsford,[3] Katsumi Ida,[4] Mikiro Yoshinuma,[4] Motoshi Goto,[4,5] Yasuko Kawamoto,[4,5] Tomoko Kawate,[8] Tokihiko Tokuzawa,[4,5] Ichihiro Yamada[4,5]

[1]Laboratorio Nacional de Fusión, CIEMAT, Madrid, Spain
[2]Max-Planck Institute for Plasma Physics, Greifswald, Germany
[3]Princeton Plasma Physics Laboratory, Princeton, NJ, USA
[4]National Institute for Fusion Science, National Institutes of Natural Science, Toki, Japan
[5]The Graduate University of Advanced Studies, SOKENDAI, Toki, Japan
[6]Interdisciplinary Graduate School of Engineering Sciences, Kyushu University, Kasuga, Japan
[7]Department of Electrical Engineering, Royal Institute of Technology, Stockholm, Sweden-
[8]National Institutes for Quantum Science and Technology, Naka Institute for Fusion Science and Technology, Naka, Japan

*Contact author: daniel.medina@ciemat.es

**ABSTRACT**. An enhancement of core impurity transport is observed in high-density plasmas of the stellarator LHD heated by neutral beam injection when continuous lithium (Li) granule dropping is performed. In the experiments reported here, in which the TESPEL is employed to inject trace amounts of titanium (Ti) and molybdenum (Mo) into the plasma core, confinement times for these impurities are seen to reduce significantly when Li dropping is applied, this reduction being more notable for Mo. In order to gain some initial insight into these observations, simulations are performed using the drift-kinetic transport code SFINCS for the Mo case. These simulations indicate that, while neoclassical transport prevails for the main plasma components (electrons, majority ions and low Z impurities), the classical contribution seems to be dominant for transporting Mo impurities. In summary, this work reports the first experimental observation of the degradation of mid-Z and high-Z impurity confinement induced by the continuous dropping of Li granules into a high-density stellarator plasma. In the case of the Mo impurity, simulations suggest that classical transport is the key mechanism underlying the enhanced impurity transport.

A crucial challenge for developing stellarator-type devices as fusion reactors is the identification of operational scenarios that ensure long bulk ion particle and energy confinement times as well as impurity accumulation avoidance, especially for high-Z elements that can induce plasma radiative collapse [1]. This is most challenging for high-density regimes in which confinement is extended due to the creation of an inwards-directed ambipolar radial electric field (ion-root) and a drop in diffusive impurity transport [2, 3]. Indeed, the reduction of anomalous transport due to plasma turbulence is a standard means to improve fuel particle confinement [4]. However, this can also enhance impurity confinement and lead to accumulation, as observed experimentally and predicted by neoclassical (NC) simulations, with the exception of the high-density H-mode in W7-AS where impurity accumulation was mitigated [1]. In the case of the Large Helical Device (LHD), enhanced impurity transport has been observed so far for a few plasma conditions only. For instance, in low-density, high-ion-temperature plasmas, an impurity hole (extremely hollow impurity profile) was observed and the role of neoclassical/anomalous transport was studied extensively [5-8]. Moreover, while Ion Temperature Gradient (ITG) turbulence induced inward-directed impurity fluxes [7, 8], detailed NC simulations revealed that thermo-diffusion plays an essential role in outward impurity fluxes. In contrast, in high-density LHD plasmas, only the use of additional ECRH has been found to pump out impurities from the core region [9, 10], albeit the physical mechanisms remain unclear.

Here, we perform TESPEL (Tracer-Encapsulated Solid Pellet) injections [11] into high-density, low-temperature plasmas of LHD with NBI-heating. These injections are made into discharges with and without continuous lithium (Li) granule deposition in the plasma edge in order to investigate the impact of granule dropping on core impurity transport. This is motivated by the fact that low-Z granule dropping has been shown to improve confinement, as observed in previous experiments using low-Z (B, BN, or C) powders dropping technique [12-14] while its effect on impurity confinement had not been investigated in any experimental device to date.

In this letter, we present experimental evidence revealing that, contrary to expectations, continuous Li-granule dropping reduces the confinement times of mid-Z and a high-Z impurities (the latter in particular) in a high-density stellarator plasma in which a negative radial electric field exists across the full plasma radius. Then in order to obtain some insight on these experimental findings, simulations made using a drift-kinetic transport code for the high-Z case are presented and discussed.

The experiments shown here are performed in the LHD, the largest superconducting heliotron-type device with a major radius of 3.9 m, a minor radius of 0.6 m and typical plasma volume of 30 $m^3$ [15]. A comprehensive collection of diagnostics is available [16]. Those of interest include Charge-Exchange recombination Spectroscopy (CXS) [17], 2-dimensional Phase Contrast Imaging (2D-PCI) [18], visible Bremsstrahlung measurement [19], as well as Vacuum and Extreme Ultraviolet (VUV/EUV) spectrometers [20]. LHD is equipped with an Impurity Powder Dropper (IPD), which allows the deposition on the plasma edge of sub-millimeter impurity grains by gravity [21]. The TESPEL, which allow the injection of known quantities of impurities at pre-determined radii and times [22, 23], is used for titanium (Ti) and molybdenum (Mo) impurity tracer injection. Experiments are conducted with the so-called inward-shifted magnetic configuration, vacuum magnetic axis position at $R_{ax}$ = 3.6 m and magnetic field strength at the magnetic axis of $B_{ax}$ = 2.75 T. Hydrogen ($H_2$) gas is used as a working gas. The heating pattern applied is 3 MW of ECRH power from 3 s to 5 s together with 3 MW of tangential NBI from 3.3 s to 5 s, with tangential NBI heating being increased to 7 MW from 5 s to 7 s and 4 MW of perpendicular modulated NBI for the CXS diagnostic. Reference plasma parameters are line-averaged density $n_{e,avg}$ ~5.3 x $10^{19}$ $m^{-3}$, central electron temperature, $T_{e,0}$, ~2.2 keV, central ion temperature, $T_{i,0}$, ~1 keV and diamagnetic energy, $W_p$, ~880 kJ. Li-granules are dropped continuously into the plasmas from ~4.8 s until discharge end (this is identified by Li I emissions (see Fig. 1e)). Comparing signals for discharges with and without Li-granule dropping, it is found that the Li modifies various plasma parameters, *e.g.*, $n_{e,avg}$ increases slightly with $\Delta n_{e,avg}/n_{e,avg}$ < 5 %, $T_{e0}$ rises where $\Delta T_{e0}/T_{e0}$ ~12.5 %, plasma stored energy, $W_p$, grows by ~ 10 % higher and $Z_{eff}$ goes up from ~1.8 to ~2.3. Moreover, the observed ~25 % reduction in Hα emission might indicate a reduction of particle fueling under almost constant line-averaged electron density. For instance, due to feedback control, $H_2$ gas puffing is reduced in the time period from 4.8 s to 6 s for discharge #189894 (with Li). However, after 6 s, the gas puff increases again to values that are slightly higher than those during reference discharge #189896. In this same time period, Hα remains moderately below the Hα level in the reference, this is attributed to possible changes in wall recycling due to Li granules. Whilst these observations would be compatible with improved bulk-ion particle confinement, a complete particle balance analysis, such as done in Ref. [24], would be needed to confirm this point. Finally, comparing signals in Fig. 1 for discharges with (#189894) and without (#189882) TESPEL, it is apparent that the effect of TESPEL on main plasma parameters is minor compared to modifications due to Li-dropping.

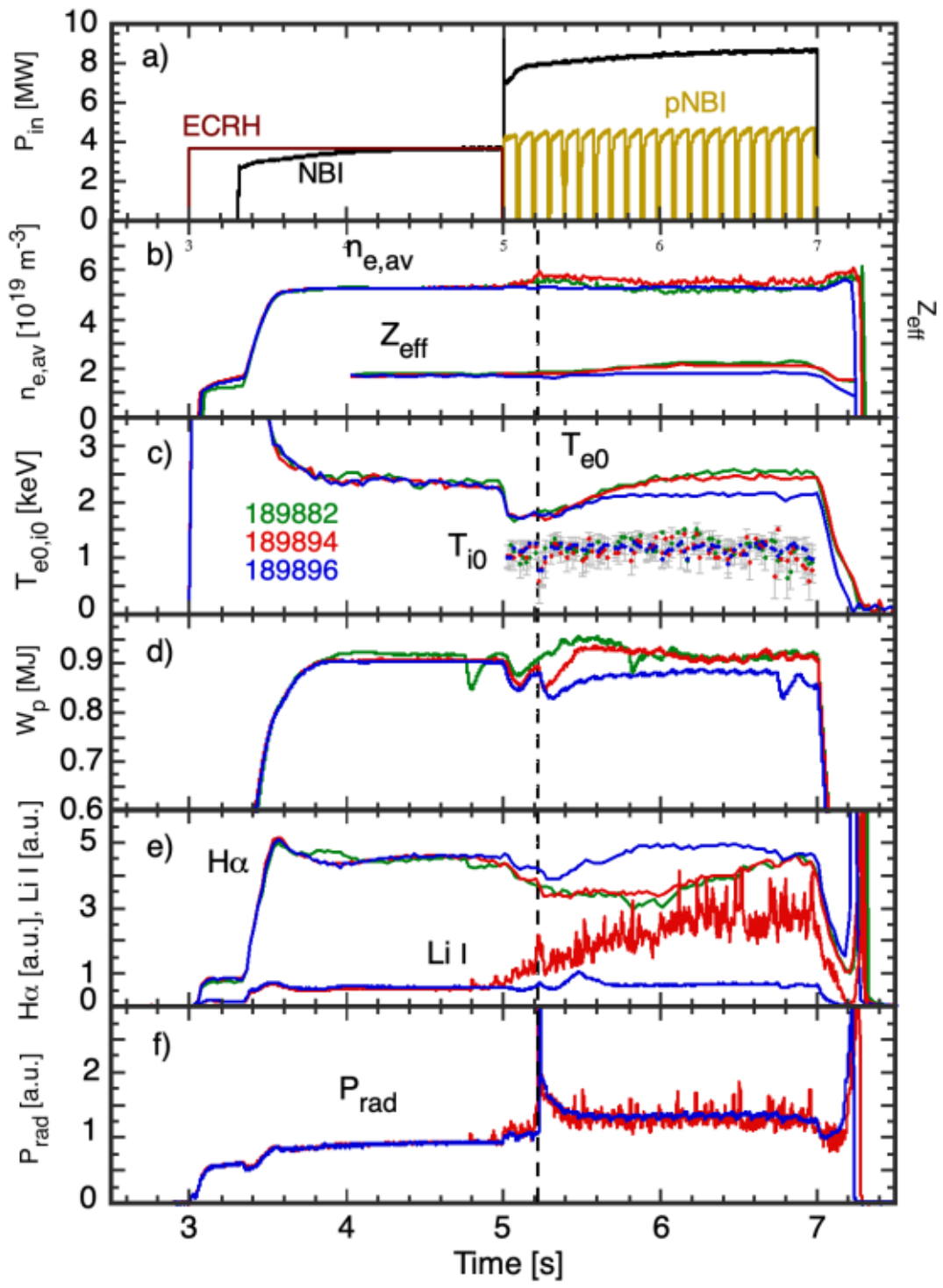

Figure 1. Time traces of a) ECRH, tangential (NBI) and perpendicular (pNBI) neutral beams, b) line-averaged electron density, $n_{e,avg}$, and $Z_{eff}$ (from 4 s to 7.2 s for clarity), c) central $T_e$ and $T_i$, d) stored plasma energy (the short dip at t ~6.8 s corresponds to a brief NBI breakdown), e) Balmer Hα (656.3 nm) and Li I (670.8 nm) emissions, and f) total radiated power, $P_{rad}$ for #189882 (Li-granules only - green), #189894 (with Li-granules and Mo-TESPEL injection - red) and #189896 (reference shot without Li-granules but with Mo-TESPEL injection - blue). Li I and $P_{rad}$ signals are not shown for #189882 to avoid figure overload. These are similar to their traces for #189894. TESPEL injections are made at 5.225 s (vertical dash - dash).

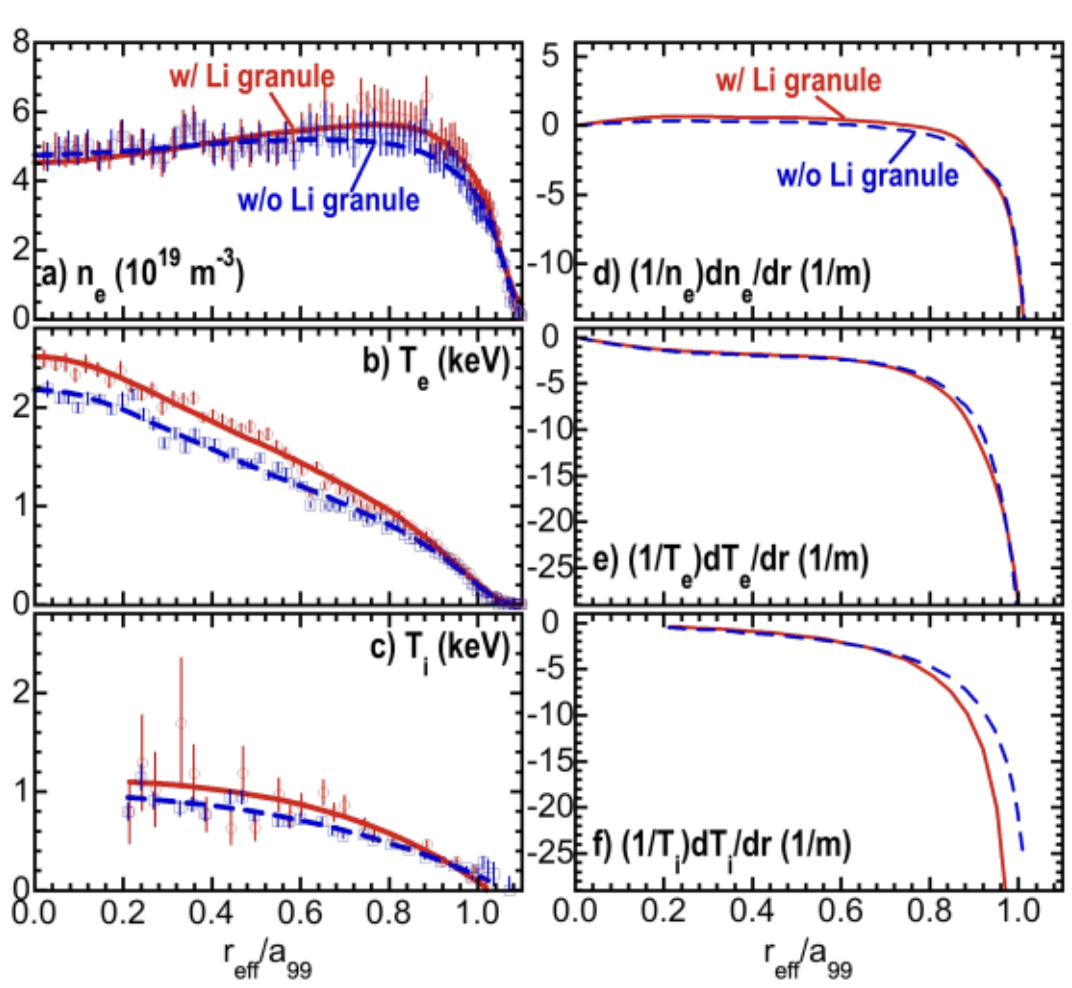

Figure 2. Radial profiles of a) electron density, b) electron temperature and c) ion temperature at t ~ 6 s for #189894 (with Li granules - red) and #189896 (without Li granules - blue). Also shown are inversed scale lengths of d) electron density, $L_{ne}^{-1}$, e) electron temperature, $L_{Te}^{-1}$, and f) ion temperature, $L_{Ti}^{-1}$, at t ~ 6 s. pNBI attenuation limits reliable $T_i$ measurements to $r_{eff}/a_{99}$ > 0.2. In both cases, a Mo-TESPEL is injected at t = 5.225 s

Nonetheless, it is found that Li-granules modify profile shapes. For instance, in Fig. 2, $n_e$ increases in the edge region (0.7 < $r_{eff}/a_{99}$ < 1) while remaining constant in the core ($r_{eff}/a_{99}$ < 0.7). Here, $r_{eff}$ and $a_{99}$ denote effective minor radius and minor radius, respectively, the latter within which 99 % of total plasma stored energy is confined. This rise in edge density leads to a slightly hollower density profile. It should be noted that the density profile does not change as dramatically as the HDH mode observed in the W7-AS stellarator. Moreover, both core $T_e$ and $T_i$ are seen to increase during Li-granule dropping, the latter despite the large error bars (due to reduced signal levels caused by CXRS beam attenuation in the plasma center). Finally, whilst the inversed scale lengths of $n_e$ and $T_e$ appear to be almost identical for both situations, the inversed scale length of $T_i$ seems to increase during Li-granule dropping at $r_{eff}/a_{99} > 0.8$.

In order to evaluate impurity confinement times with/without Li-granules, an impurity is injected with TESPEL at 5.225 s, as indicated by a dashed vertical line in Fig. 1. Ti (Z = 22) and Mo (Z = 42) impurities are injected to identify possible Z dependence of impurity transport. Impurity particle amounts are: 2.94 x $10^{17}$ Ti atoms for #189891, 2.82 x $10^{17}$ Ti atoms for #189893, 4.98 x $10^{17}$ Mo atoms for #189894, and 5.92 x $10^{17}$ Mo atoms for #189896. The radial deposition location of impurities injected by TESPEL is estimated to be around $r_{eff}/a_{99} = 0.75$ from time-of-flight measurements, this being inside the density profile shoulder. Outward drifting of the ablated tracer cloud, as seen for cryogenic pellet clouds, is known to be negligible for TESPEL [25], thus it can be assumed that the full tracer amount is deposited at that radius. In addition, TESPEL injection causes a small transient rise in $n_{e,avg}$, a 5 % decrease in plasma stored energy that recovers fully after 200 ms, and a 50 % reduction in $T_i$ that recovers its pre-injection level after ~50 ms. Next, Figure 3 shows the temporal evolutions of Ti XX (25.93 nm) and Mo XXXII (12.79 nm) spectral lines measured with the EUV/VUV spectrometer, SOXMOS [26]. These lines are chosen for determining impurity confinement times as $Ti^{19+}$ and $Mo^{31+}$ are the most highly ionized states for these plasmas. Hence, the data points in Fig. 3 are obtained from the measured intensities of these emission lines, and the corresponding confinement times are determined by exponential fits to these points in the decay phase. In this way, the Ti confinement time without Li granules is found to be 1.38 ± 0.18 s, while that with Li granules is 1.15 ± 0.4 s (~17 % reduction). Here a time window from 5.475 to 6.525s is used to fit the data points, as an NBI breakdown occurred in discharge #189891 from 6.55 s. It should be noted that the confinement time for Ti without Li granules is consistent with that shown in Fig. 1 of Ref. [3] for the same heating power (7 MW of NBI). Next, the confinement time for Mo without Li granules is 6.34 ± 26.3 s and with Li granules is 1.43 ± 0.8 s (~78 % reduction). In this case, a time window from 6.075 to 7.025s was used as, due to the higher ionization energy of $Mo^{31+}$, the rise of the line emission takes longer than for $Ti^{19+}$ and thus the decay phase starts at a later time. It must be noted here that the large error estimated for Mo in the reference case is due to the short time window compared to the estimated decay time, however the faster decay in presence of Li is clear by the evolution of the data points in both cases. Comparing these times, it would appear that the higher the Z impurity, the more the impurity confinement time is reduced in plasma by Li granules.

For these plasmas, simulations are performed using the STRAHL impurity transport code [27, 28] to compare the time evolution of the emission lines observed spectroscopically with simulation estimations of impurity transport coefficients that closely reproduce the experimental results. The spectroscopic measurements used here integrates light along the major radius, thus, we assume that the diffusion coefficient is spatially uniform. Thus, the solid lines in Fig. 3 show the STRAHL results (continuous lines) which reproduce well the decay phase of the impurity line emissions (data points), where impurity confinement degradation (difference with and without Li) is quite prominent. The temporal evolution of Ti XX without Li granules is well reproduced with D = 0.11 $m^2/s$ and the convection velocity, V, profile (minimum V of -1.55 m/s) shown by the sub-plot in Fig. 3a). For the temporal evolution of Ti XX with Li granules, the same D and a slightly narrower negative V profile (minimum V of -0.95 m/s) can reproduce the experimental results. We also find that the temporal evolutions of Mo XXXII are also well reproduced with D = 0.085 m^2/s, $V_{min}$ = -2.0 m/s (without Li) and $V_{min}$ = -1.2 m/s (with Li). In this case, the region where V becomes negative in the transport of Mo XXXII is the same as in the transport of Ti XX.

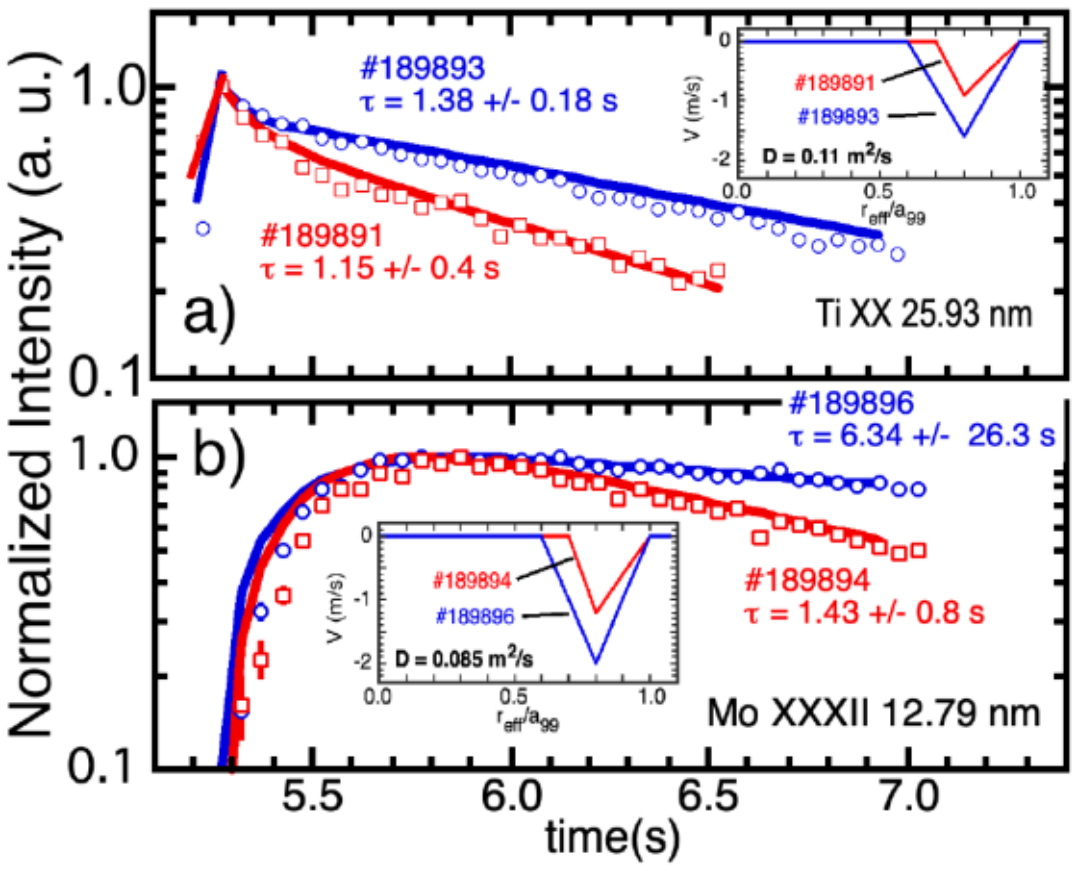

Figure 3. Semi-log plots showing normalized intensities of a) Ti XX and b) Mo XXXII emissions lines (open squares) for #189893 and #189896 (without Li-granules - blue) and for #189891 and #189894 (with Li-granules - red). TESPELs are injected at 5.225 s. Solid lines are STRAHL code predictions for these injections. Inserted plots show convective velocity (V) profiles and diffusion coefficient values (D) used for STRAHL. Impurity confinement times (decay times), τ, are obtained by fits to experimental data. Uncertainty bars for data points and decay times are included.

Therefore, this might suggest that the impact region of Li granule dropping on core impurity transport is the same for both Ti and Mo impurities.

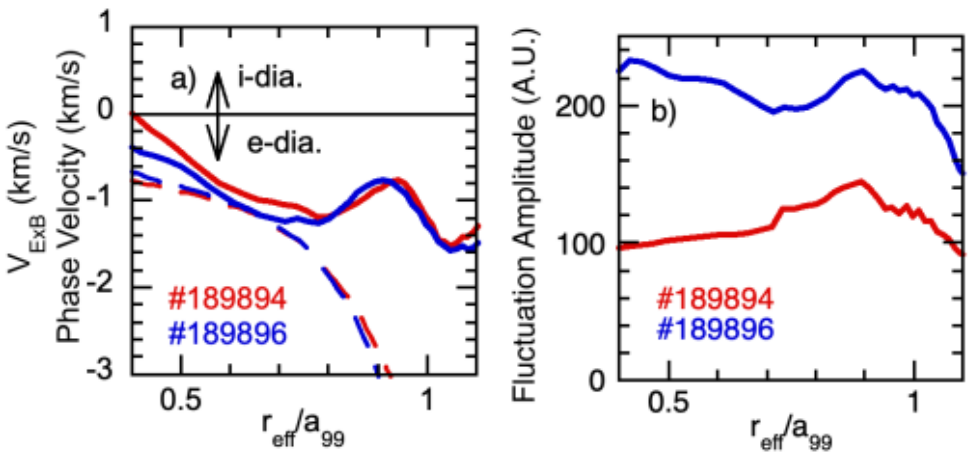

Figure 4. Radial profiles of a) fluctuation phase velocity (solid) at 6 s measured with 2D-PCI and $V_{ExB}$ velocities estimated by SFINCS (dash - dash) and b) fluctuation amplitudes for #189894 (with Li-granule dropping - red) and #189896 (reference, without Li-granules - blue).

Now, in order to understand whether Li-granule dropping can modify impurity transport in the range of $r_{eff}/a_{99} > 0.6$ it is first necessary to consider the role of the radial electric field, $E_r$. This can be estimated by the 2D-PCI diagnostic, which measures ion scale turbulence where k = 0.1 - 0.8 $mm^{-1}$, f = 20 - 500 kHz. For instance, in Fig. 4a), the fluctuation phase velocities are dominated by the poloidal components, thus they are considered to be close to ExB poloidal rotation velocities. The $E_r$ inferred from the phase velocities is negative and nearly identical with and without Li-granules. In the same figure, the $V_{ExB}$ predicted by the collisional transport code SFINCS [29] is also similar to the phase velocities measured with the 2D-PCI in the range of $0.4 < r_{eff}/a_{99} < 0.7$ for both cases. Consequently, the inferred $E_r$ remains negative and exhibits similar values both with and without Li-granules, thus variations in $E_r$ cannot explain the enhanced impurity transport observed.

In parallel, as shown in Fig. 4b, fluctuation amplitudes measured with 2D-PCI are reduced significantly in plasmas with Li granule dropping with respect to the reference. It is considered here that the dominant anomalous transport here is more likely to be Resistive Interchange (RI) turbulence [30], rather than ITG which is presumably stable, *i.e.*, the threshold for ITG is $L_{Ti}^{-1}$ ~2.5 $m^{-1}$ [31]. The lower growth rate of RI with Li-granules could be associated with the higher $T_e$, higher $Z_{eff}$ (and lower pressure gradient) as reported in previous hydrogen/deuterium isotope experiments [31]. Thus, while reduced RI turbulence can provide a possible explanation for improved energy confinement, it is unlikely to favor impurity removal, in fact, the opposite would be expected. Hence, a different physical mechanism must be contemplated in order to explain these experimental observations in reduced-turbulence regime due to Li-granule edge plasma deposition.

Finally, classical and neo-classical particle fluxes are computed using SFINCS for discharges with/without Li-granule dropping and with Mo TESPEL injection, as a larger reduction of impurity confinement time is observed for the Mo impurity case. SFINCS simulates particle, momentum and heat flows by solving the drift-kinetic equation. As shown in Fig. 5, total electron and proton fluxes are dominated by the NC contribution and they increase in the case of Li-granule dropping (#189894).

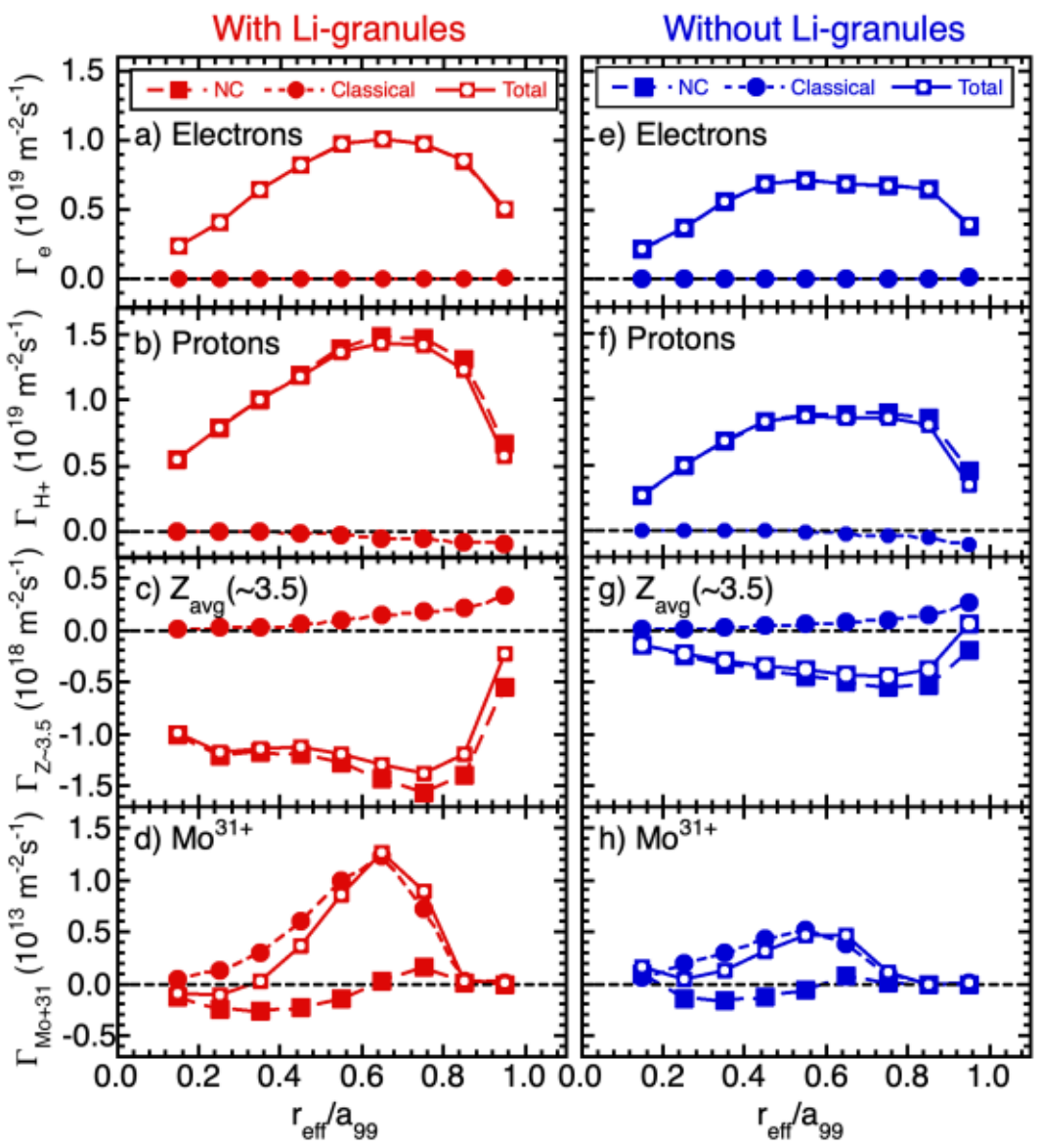

Figure 5. NC (squares, long dash-dash), classical (dots, short dash-dash) and total (open circles contained in squares, continuous) particle fluxes normalized to pseudo-densities for a) electrons, b) $H^+$ ion, c) effective impurity content with $Z_{avg}$ = ~3.5 and d) trace $Mo^{+31}$ ions as a function of normalized minor radius. Predictions are made with the SFINCS code for #189894 (left - red, with Li-granule dropping) and #189896 (right - blue, without Li-granule dropping).

In the simulations, an “average” impurity ion with average charge equal to ~3.5 is considered to account for low-Z impurities, and, for simplicity, the averaged impurity density profile is assumed to be the same as the electron density. This reduces the cost of simulations since the system being solved grows with the number of species and the results are substantially the same in terms of $E_r$ and bulk particle fluxes. It is found from these simulations that NC fluxes, which dominate for the "averaged impurity", become more negative at all radii for the with Li case in order to maintain ambipolarity. See Fig. 5. Indeed, such enhanced inward-directed NC fluxes may increase the impurity density gradient in the plasma core, thus reducing core turbulence, as postulated recently [32].

In high-density LHD plasmas, it is considered that a radial electric field, $E_r$, which is negative, drives an inward-directed impurity particle flux, a situation that can lead to impurity accumulation. Nonetheless, looking at the Mo XXXII line decay for the reference discharge of Fig. 3, it is apparent that $Mo^{31+}$ ions are expelled, albeit slowly, from such plasma. This indicates that an outward-directed particle flux cancels the inward particle flux arising from such a negative $E_r$. Indeed, in the simulations from SFINCS presented in Fig. 5h for $Mo^{31+}$ ions without Li-granule dropping, predictions show that the total $Mo^{31+}$ particle flux, which is small and directed outwards at $r_{eff/a99} = 0.8$, is supported mainly by the classical channel. However, in Fig. 5d, the total $Mo^{+31}$ particle flux is outward enhanced by Li-granule dropping, peaking at $r_{eff/a99} = 0.65$, this being due to the strong increase in its classical channel. In contrast, the $Mo^{+31}$ particle flux due to NC transport remains small and similar for both situations. It should be mentioned here that coronal equilibrium is assumed for the $Mo^{31+}$ ion density profile in these simulations and that its level is set to a trace value to avoid significantly altering $Z_{avg}$. Also, it should be noted that similar results are obtained when a Gaussian profile of $Mo^{+31}$ ion density, centered around its expected deposition position, is assumed. Hence, from these findings it is interpreted that an enhancement of the collisional transport channel arises from collisions between Mo ions (the same amounts are injected for both situations) and Li ions (when dropped). Thus, classical processes contribute strongly to high-Z impurity transport in the LHD plasmas studied here when Li-granules are dropped, while NC processes dominate the transport of low-Z impurities and bulk ions for both with and without Li-granule dropping cases. Now, returning to Fig. 3, an increase in classical outward flux could explain also the reduction in the inward pinch (negative convection velocity) predicted from STRAHL in the Li-dropping case.

In summary, continuous Li-granule dropping into high-density LHD plasmas when heated by NBI only results in enhanced transport of mid- and high-Z tracer impurities injected into in the core of such plasmas. Simulations performed here using the drift-kinetic transport code SFINCS indicate that the classical transport channel for high-Z impurities is enhanced in the region $r_{eff}/a_{99} > 0.6$ when Li-granules are dropped into the plasm edge, this being more likely due to increased collisionality between Ti/Mo and Li ions, which leads to stronger outward transport. Indeed, the importance of the classical transport channel for high-Z impurities has been pointed out analytically in optimized stellarators, such as W7-X [33]. Thus, this letter provides first experimental evidence that, even in non-optimized stellarators such as LHD, the transport of mid/high-Z impurities can be enhanced by a strong contribution from the classical transport channel when Li granules are continuously dropped. A condition for classical transport channels to become important, even in unoptimized stellarators, is that competing NC transport channels are small.

These findings may be applicable to other magnetic confinement devices, including tokamaks other than the LHD, and may open new route for impurity control. Currently, impurity powder droppers have been installed in various magnetic confinement devices, and experimental verification of these findings is anticipated. The key method is to create a state in which classical transport exceeds neoclassical transport. The future challenge is to find the optimal conditions under which the impurity powder dropper can be operated in accordance with its original purpose, real-time wall conditioning, while still achieving sufficient impurity control. Since the installation of an impurity powder dropper is currently planned for ITER [34], experimental verification of the device in existing devices is an extremely important and urgent task.

## ACKNOWLEDGMENTS

This work has been carried out within the framework of the EUROfusion Consortium, funded by the European Union via the Euratom Research and Training Programme (Grant Agreement No 101052200 - EUROfusion). Views and opinions expressed are however those of the author(s) only and do not necessarily reflect those of the European Union or the European Commission. Neither the European Union nor the European Commission can be held responsible for them. It is partially financed by grants PID2020-116599RB-I00 and PID2023-148697OB-I00 funded by MCIN/AEI/ 10.13039/501100011-033 and by ERDF, A Way of Making Europe. It is partially supported also by the U.S. DOE under Contract No. DE-AC02-09CH11466 with Princeton University, by JSPS KAKENHI JP23KK0054 and by NIFS grant administrative budgets (10203010LHD105 and 10201010PSU003). Computing resources were provided on the computer Stellar operated by the Princeton Plasma Physics Laboratory and Princeton Institute for Computational Science and Engineering.

[1] H. Maaßberg, C. D. Beidler and E. E. Simmet, Density control problems in large stellarators with neoclassical transport, Plasma Phys. Control. Fusion 41, 1135 (1999).
[2] S. Sudo, A review of impurity transport characteristics in the LHD, Plasma Phys. Control. Fusion 58, 043001 (2016)
[3] R. Burhenn et al., On impurity handling in high performance stellarator/heliotron plasmas, Nucl. Fusion 49, 065005 (2009).
[4] K. Tanaka et al., Particle Transport of LHD, Fusion Sci. Tech. 58, 70 (2010).
[5] K. Ida et al., Observation of an impurity hole in a plasma with an ion internal transport barrier in the Large Helical Device, Phys. Plasmas 16, 056111 (2009).
[6] K. Tanaka et al., Turbulence Response in the High Ti Discharge of the LHD, Plasma Fusion Res. 5, S2053 (2010).
[7] D. R. Mikkelsen et al., Quasilinear carbon transport in an impurity hole plasma in LHD, Phys. Plasmas 21, 082302 (2014).
[8] M. Nunami et al., Gyrokinetic simulations for turbulent transport of multi-ion-species plasmas in helical systems, Phys. Plasmas 27, 052501 (2020).
[9] N. Tamura et al., Mitigation of the tracer impurity accumulation by EC heating in the LHD, Plasma Phys. Control. Fusion 58, 114003 (2016).
[10] N. Tamura et al., Observation of the ECH effect on the impurity accumulation in the LHD, Phys. Plasmas 24, 056118 (2017).
[11] S. Sudo and N. Tamura, Tracer-encapsulated solid pellet injection system, Rev. Sci. Instrum. 83, 023503 (2012).
[12] F. Nespoli et al., Observation of a reduced-turbulence regime with boron powder injection in a stellarator, Nat. Phys. 18, 350 (2022).
[13] R. Lunsford et al., Real-time wall conditioning and recycling modification utilizing boron and boron nitride powder injections into the Large Helical Device, Nucl. Fusion 62, 086021 (2022).
[14] F. Nespoli et al., A reduced-turbulence regime in the Large Helical Device upon injection of low-Z materials powders, Nucl. Fusion 63, 076001 (2023).
[15] A. Iiyoshi et al., An Overview of the Large Helical Device Project, Fusion Tech. 17, 169 (1990).
[16] K. Kawahata et al., Overview of LHD Plasma Diagnostics, Fusion Sci. Tech. 58, 331 (2010).
[17] M. Yoshinuma et al., Charge-Exchange Spectroscopy with Pitch-Controlled Double-Slit Fiber Bundle on LHD, Fusion Sci. Tech. 58, 375 (2017).
[18] K. Tanaka et al., Two-dimensional phase contrast imaging for local turbulence measurements in large helical device, Rev. Sci. Instrum. 79, 10E702 (2008).
[19] Y. Kawamoto et al., Qualitative Improvement of Zeff Diagnostic Based on Visible Bremsstrahlung Profile Measurement by Changing Observation Cross Section in LHD, Plasma Fusion Res. 16, 2402072 (2021).
[20] M. Goto et al., Passive Spectroscopy in Visible, VUV, and X-ray Ranges for LHD Diagnostics, Fusion Sci. Tech. 58, 394 (2010).
[21] A. Nagy et al., A multi-species powder dropper for magnetic fusion applications, Rev. Sci. Instrum. 89, 10K121 (2018).
[22] N. Tamura et al., Tracer-Encapsulated Solid Pellet (TESPEL) injection system for the TJ-II stellarator, Rev. Sci. Instrum. 87, 11D619 (2016).
[23] R. Bussiahn et al., Commissioning of the tracer-encapsulated solid pellet (TESPEL) injection system for Wendelstein 7-X and first results for OP1.2b, Plasma Phys. Control. Fusion 66, 115020 (2024).
[24] J. Miyazawa et al., Role of recycling flux in gas fuelling in the Large Helical Device, J. Nucl. Mat. 313-316, 534 (2003).
[25] K. J. McCarthy et al., Comparison of cryogenic (hydrogen) and TESPEL (polystyrene) pellet particle deposition in a magnetically confined plasma, EPL 120 25001 (2017).
[26] J. L. Schwob et al., High-resolution duo-multichannel soft x-ray spectrometer for tokamak plasma diagnostics, Rev. Sci. Intrum. 58, 1601 (1987).
[27] K. Berhinger, Description of the impurity transport code 'STRAHL', JET Report, JET-R (87) 08 (1987).
[28] R. Dux STRAHL user manual (2006).
[29] M. Landreman, H. M. Smith, A. Mollén, and P. Helander, Comparison of particle trajectories and collision operators for collisional transport in nonaxisymmetric plasmas, Phys. Plasmas 21, 042503 (2014).
[30] T. Kinoshita et al., Turbulence Transition in Magnetically Confined Hydrogen and

Deuterium Plasma, Phys. Rev. Lett. 132, 235101 (2024).
[31] K. Tanaka et al., Collisionality dependence and ion species effects on heat transport in He and H plasma, and the role of ion scale turbulence in LHD, Nucl. Fusion 57, 116005 (2017).
[32] J. M García-Regaña et al., Reduction or Enhancement of Stellarator Turbulence by Impurities, Phys. Rev. Lett. 133, 105101 (2024).
[33] S. Buller et al., The importance of the classical channel in the impurity transport of optimized stellarators, J. Plasma Phys. 85, 175850401 (2019).
[34] J. Snipes et al., Initial design concepts for solid boron injection in ITER, Nucl. Mat. and Energy 41, 101809 (2024).